\documentclass[11pt,twoside]{article}
\usepackage{./asp2014}
\usepackage{graphicx}
\usepackage{wrapfig}
\usepackage{lscape}
\usepackage{rotating}
\usepackage{epstopdf}

\aspSuppressVolSlug
\resetcounters

\begin{document}

\title{NGVLA Memo 57: Imaging the Distribution of Solids in Planet-forming Disks undergoing Hydrodynamical Instabilities with the Next Generation Very Large Array}

\author{L. Ricci$^1$, M. Flock,$^2$ D. Blanco$^1$, and W. Lyra$^1$}
\affil{$^1$ Department of Physics and Astronomy, California State University Northridge,
18111 Nordhoff Street, Northridge, CA 91330, USA \email{luca.ricci@csun.edu}}
\affil{$^2$ Max-Planck Institute for Astronomy, K{\"o}nigstuhl 17, 69117, Heidelberg, Germany}

\paperauthor{L. Ricci}{luca.ricci@csun.edu}{}{California State University Northridge}{}{Northridge}{CA}{91330}{USA}
\paperauthor{M. Flock}{}{}{National Radio Astronomy Observatory}{}{Socorro}{NM}{87801}{USA}
\paperauthor{D. Blanco}{}{}{California State University Northridge}{}{Northridge}{CA}{91330}{USA}
\paperauthor{W. Lyra}{}{}{California State University Northridge}{}{Northridge}{CA}{91330}{USA}


\begin{abstract}
We present simulations of the capabilities of the Next Generation Very
Large Array to image at high angular resolution substructures in the dust emission of protoplanetary disks. 
The main goal of this study is to investigate the kinds of substructures that are expected by state-of-the-art 3D simulations of disks and that an instrument like the ngVLA, with its current design, can detect.  The disk simulations adopted in this investigation consist of global 3D radiation-hydrodynamics models with embedded particles, the latter representing dust grains. Our work shows that the ngVLA can detect and spatially resolve, down to sub-astronomical unit scales in disks in nearby star forming regions, the dust continuum emission at 3mm from azimuthal asymmetric structures, as well as from weak rings and gaps produced in these models as a consequence of the vertical shear instability (VSI). This hydrodynamical instability has been proposed to generate turbulence in regions of weak coupling between the disk gas and magnetic field, as well as to form vortices which may be preferred locations of planetesimal formation.

\end{abstract}

\section{Introduction}

The discovery of thousands of exoplanets over the last couple of decades has shown that
the birth of planets is a very efficient process in nature \citep[e.g.,][]{Burke:2015}. However, the physical mechanisms responsible for their origin are often poorly understood, and observational evidence for several of the proposed mechanisms is lacking. The currently accepted theory is that
planets form in young circumstellar disks through the agglomeration of small dust particles into km-sized
``planetesimals'', which are massive enough to gravitationally attract gas and other solids
in the disk.

The formation of planetesimals is one of the most critical problems for theories of planet formation.
In the simplest assumption of a gas-rich disk with density and temperature decreasing
further from the star, small solids radially drift towards the star as a consequence of the aerodynamical
drag by the gas rotating at sub-Keplerian speeds
\citep{Weidenschilling:1977}. Models
of solids evolution in disks have calculated radial drift timescales which are too short to allow them to grow to planetesimals. Moreover, the associated high velocities
correspond to kinetic energies much higher than the binding energies of small solids as
measured in laboratory experiments \citep{Testi:2014}, so that collisions lead to destruction
instead of growth.
Inward radial drift can be slowed down or stopped if there are local over-densities in the gas
capable of trapping particles. In these over-densities the dust-to-gas ratio may increase up to
the point at which the dusty layer becomes unstable and fragment. These fragments might
eventually gravitationally collapse forming planetesimals. The origin of these over-densities are
likely dynamical gas instabilities \citep{Lyra:2019}, such as the Rossby wave instability \citep{Lovelace:1999,Li:2001}, convective overstability \citep{Klahr:2014,Lyra:2014}, vertical shear instability 
\citep[][]{Nelson:2013}, zombie vortex instability \citep{Marcus:2015}, as well as streaming instabilities, in which the dust is an active ingredient in generating the density fluctuations \citep{Youdin:2005,Johansen:2007,Youdin:2007}. 
A natural prediction of these models is the presence of strong local accumulations of small solids
in the disk \citep[e.g.,][]{Whipple:1972,Haghighipour:2002,Barge:1995,Lyra:2009,Chiang:2010,Johansen:2014}.

In the last years, ALMA has found striking evidence for these local accumulations of small
solids at distances $> 10-20$ au from the star \citep[e.g.][]{vanderMarel:2013}. The results of the ALMA
observations demonstrate that some of the instabilities invoked by theory to explain the origin of planets are occurring in real disks, but they are limited to the outskirts of the planet-forming
regions of disks (in the Solar System, $\approx 30-50$ au corresponds to the Kuiper Belt). Furthermore,
models predict that solids with sizes of $\sim 1-10$ mm should show significantly higher levels of
concentration than smaller dust grains. Constraining their spatial distribution is therefore key to inform and test the models of planetesimals formation. Solids with these sizes can be best traced by mapping
the dust thermal emission at wavelengths longer than 1 mm.

Given its very high sensitivity and angular resolution at long wavelengths \citep{Murphy:2018}, the ``Next Generation Very Large Array'' (ngVLA) will
have the right technical capabilities to reveal and investigate the morphology of the birthsites of
planetesimals in the planet-forming regions of disks, i.e. $< 10-30$ au from the star. The goal of
this project is to study in a quantitative way the potential of the ngVLA to image the birthsites
of planetesimals in nearby planet-forming disks. To derive predictions for models of the early evolution of solids in young disks, we performed state-of-the-art global 3D hydrodynamical radiative simulations of disks undergoing vertical shear instabilities based on the method described in \citet{Flock:2017}.

\section{Methods: from disk simulations to ngVLA and ALMA observations}

We describe here the method used to simulate observations with the ngVLA and ALMA for the dust continuum emission of disks undergoing vertical shear instabilities.

\subsection{Disk model}

In order to derive predictions for the appearance of disk models with particle trapping driven by hydrodynamical instabilities, we perform 3D radiation hydrodynamical global simulations, in which the radiation hydrodynamic equations are solved using the hybrid flux-limited diffusion and irradiation method developed
by \citet{Flock:2013}.
This model follows the setup described in \citet{Flock:2017}. The only change is the domain size in azimuth which we extended to the full 2$\pi$ value, so that the model described here represents the first global simulation of this kind. This gives a total grid size of $1024 \times 512 \times 2044$ in the $(r, \theta, \phi)$ domain. This domain covers from 20 to 100 au in stellocentric radii and $\pm$ 0.35 in $z/r$, where $z$ is the cylindrical height across the disk.

Our disk model includes radiative transfer with stellar radiation.
The initial conditions assumed for this simulation are taken from Table 1 in \citet{Flock:2017}. 
Particles with sizes of 0.1 and 1mm are randomly distributed in the domain after the vertical shear instability has reached a quasi steady state \citep[see][]{Flock:2017}. Our simulation includes a total of 1 million particles. After a couple of dynamical times the grains start to drift radially inward towards the central star whereas they are concentrated in regions of high pressure. Moreover we find the appearance of a vortex in this kind of simulations as it was recently found by \citet{manger:2018}. Such a vortex is able to concentrate the small solids in the disk. Furthermore, the deviations in the radial pressure profile cause ``traffic jams'' in the radial drift of grains which become visible as ring-like features. 
More details on these simulations will be discussed in a future paper (Flock et al. 2019, in prep.). The main products of this simulation for the study presented here are synthetic maps for the dust continuum emission at a wavelength of 3mm.  
The synthetic model image at 3mm is displayed on the top row of Figure 1.

\subsection{ngVLA and ALMA observations}

The synthetic map obtained following the method outlined in the previous section is converted into predictions for future observations with the ngVLA and ALMA using the CASA software package \citep{McMullin:2007}.
The adopted procedure is the same as in \citet{Ricci:2018}, in which the ngVLA simulations are performed using the \texttt{SIMOBSERVE} task to generate the visibility dataset in the $(u,v)$ Fourier space, and the \texttt{SIMNOISE} task to add the noise by corrupting the visibilities.

For the ngVLA simulations we considered the original ngVLA Rev B array configuration
distributed across the US Southwest and Mexico. The Rev B configuration includes 214 antennas of 18 meter diameter, to baselines up
to 1000\,km \citep{Selina:2018}. We instead did not include the 30 additional antennas distributed to baselines up to 9000\,km, as no signal would be detectable on these very long baselines from the sources considered here. For the array configuration of the ALMA observations, we used the \texttt{alma.out28.cfg} antenna position file available in the
CASA package, which contains the longest 16 km baselines in the ALMA array.

For the imaging of the visibilities we employed the \texttt{CLEAN} algorithm with Briggs weighting, and adjusted the robust parameter and $u,v$-taper to give a reasonable synthesized beam and noise performance.
In particular, the ALMA images were computed
with a Briggs weighting with robust parameter $R = −2$ (uniform weighting), while for the ngVLA we chose
$R = − 1$. We also employed a multiscale clean approach to better recover compact emission at both high brightness and large diffuse
structures in the model.
The disk center was located at a declination of +24.0 degrees for the ngVLA simulations and -24.0 degrees for the ALMA simulations, and the assumed distance is 140pc.

\begin{figure}[t!]
\begin{center}
\includegraphics[scale=0.7]{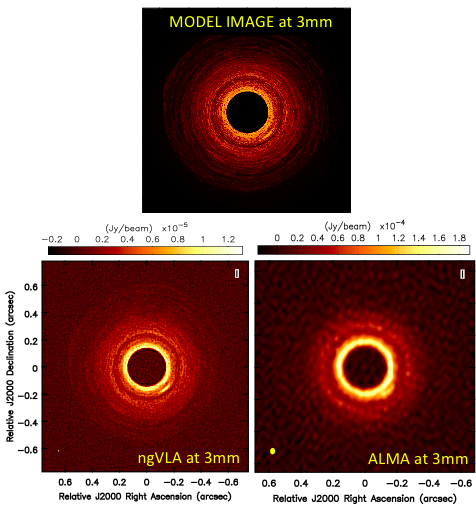}
\end{center}
\caption{\footnotesize {Images of the dust continuum emission at 3mm for our disk model, with a total flux density of 16 mJy at 3mm. Top: the model synthetic image extracted with the method described in Section 2.1.
Bottom left: The ngVLA image at 8mas $\times$ 6mas resolution. This was obtained after applying an outer taper of 5mas when imaging the visibilities using the method described in Section 2.2. The rms noise is 0.42\,$\mu$Jy\,beam$^{-1}$ and the peak signal-to-noise on the map is $\approx 50$.
Bottom right: The ALMA image at 40mas $\times$ 31mas resolution (no outer taper),
for a 8 hour synthesis.  The rms noise is 5.2\,$\mu$Jy\,beam$^{-1}$. In both images, the synthesized beam is shown as a yellow ellipse in the bottom left corner.
}}
\end{figure}

\section{Results}

\subsection{Disk Images at 3mm}

Figure~1 shows the results of our dust continuum simulations at a wavelength of 3mm for both the ngVLA and ALMA (bottom left and bottom right panels, respectively). 

For imaging of the ngVLA visibility dataset, we applied an outer taper of 5mas to increase the signal-to-noise on the map. The resulting synthesized beam has a FWHM $=$ 8\,mas $\times$ 6\,mas, corresponding to a spatial resolution of about 1.1\,au $\times$ 0.8\,au at the assumed distance of 140pc for our disk model.
The rms noise on the ngVLA map is 0.42\,$\mu$Jy\,beam$^{-1}$, which, according to the results of the study on the ngVLA sensitivity presented in \citet{Carilli:2018}, should be achieved in about 10 hours of on-source time given the current reference design for the ngVLA interferometer. The peak signal-to-noise on the ngVLA map is about 50.

The ALMA image was obtained by simulating a 8-hour long synthesis with the most extended array configuration available for ALMA in CASA. The peak signal-to-noise on the ALMA map is slightly above 30, a factor of $\approx 1.4\times$ lower than in the ngVLA image. More than the surface brightness sensitivity, the striking difference between the ngVLA and ALMA simulated observations is the angular resolution, as the ngVLA synthesized beam is smaller than the ALMA beam by a factor of 5 (in linear size). 
An even higher angular resolution could be achieved with the ngVLA with a more modest outer taper of the visibilities but at the expense of a lower signal-to-noise. For example, with a synthesized beam a factor of 2 smaller than shown in Fig. 1, the peak signal-to-noise would be decreased to values slightly above 10, high enough to characterize the brightest disk regions at a spatial resolution of 0.5 au. 

The impact of the higher resolution provided by the ngVLA is evident when one compares the ngVLA and ALMA images with the synthetic model image in Figure 1.
For example, the ALMA observations do not resolve the inner ring/gap structure, as shown by the fact the the ring appears as significantly narrower than in the ngVLA map\footnote{Note that the presence of the large inner cavity in the disk images shown in Figures 1 and 2 is an effect of the inner disk radius of 20 au considered in the disk simulation presented here.}. The better sensitivity of the ngVLA at these wavelengths also allows for a better detection of the dust emission across the whole disk, up to the fainter regions in the disk outskirts. 

Furthermore, the local concentration of dust which is visible in the south-west side of the disk is well separated and spatially resolved by the ngVLA, whereas it is not resolved along the radial direction, and not even well separated by the inner ring-like structure, by the ALMA observations. Other asymmetric structures are detected by the ngVLA, although at lower signal-to-noise levels, in the fainter disk outskirts, but are not visible in the ALMA map.

Right panel of Figure 2 shows a closer look into the area surrounding the most prominent local dust concentration, with the synthesized beam represented by the yellow ellipse with a linear size of $\approx 1$ au. This image highlights the details in the radial and azimuthal structure predicted by our disk model for the distribution of solids that the ngVLA would be capable of capturing.

\begin{figure}[t!]
\begin{center}
\includegraphics[scale=0.65]{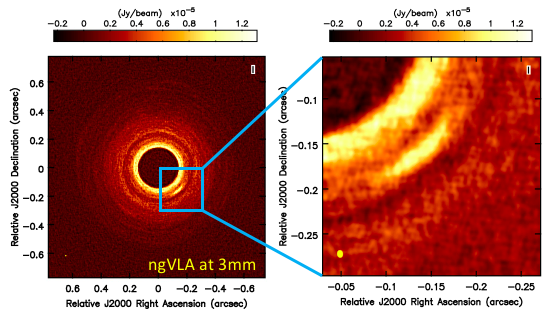}
\end{center}
\caption{\footnotesize {Closer look into the ngVLA 3mm image around the dust-trapping vortex. The ngVLA image shown here is the same as in the left panel of Figure 1. The size of the synthesized beam, shown as the yellow ellipse on the bottom left corner, corresponds to a physical scale of about 1 au at the distance of the disk. 
}}
\end{figure}

The presence of the hydrodynamical instabilities generated in the gaseous component of the disk strongly affects the dynamics of the solids, making them migrate away from their initial orbits. As a consequence of this dynamics, a low-contrast narrow gap is opened at $\approx 0.23$arcsec from the star, a bright azimuthally symmetric structure is detected at $\approx 0.15$arcsec from the star, and an asymmetric structure with a local enhancement by an order of magnitude in the dust density is spatially resolved, as seen towards the center of the right panel in Fig. 2. This strong local enhancement in the solid component is due to a vortex in the gas. This vortex-induced ``dust trap'' is an example of what is often invoked by models of the early evolution of solids in disks to trigger the formation of planetesimals. The current specifications for the ngVLA would have the imaging capabilities to spatially resolve these birthsites of planetesimals, and test the model predictions regarding their morphology and location in nearby bright disks.

\section{Conclusions}

The ngVLA will transform the field of protoplanetary disks,
allowing for imaging of their physical structure at sub-au resolution in nearby star forming regions. 

In this study we tested the imaging capabilities of the ngVLA for characterizing regions of local enhancement in the dust density. According to the models for the early evolution of solids in disks, such as the model presented in this work, these regions are necessary to concentrate small solids to higher and higher density, which may be necessary to form km-sized planetesimals, i.e. the building blocks of planets. High angular resolution observations of nearby bright disks with the ngVLA would be able to detect and characterize their morphology, e.g. their contrast and aspect ratio, and location in the disk. Observations of this kind would be key to test the predictions of physical models for the formation of vortices in the disk, as the aspect ratio of the observed structure is directly related to the vortex strength.  

Multi-wavelength observations both within the range of frequencies covered by the ngVLA, and also in combination with observations at higher frequencies with ALMA, would allow us to map the spectral index in these regions of local dust concentration. This is necessary to constrain observationally the spatial segregation of grains with different sizes as well as to better quantify the dust optical depth at these wavelengths.   

Moreover, the expected high astrometric precision of the ngVLA would allow for the investigation of the orbital motion of these asymmetric structure, as well as for possible variation in the morphology of these unstable regions on short timescales \citep[see for example the discussion in][for the case of disk substructures induced by the interaction with a planet]{Ricci:2018}. An investigation of the time-evolution of these structures will be presented in Ricci et al. (in prep.).

This work shows the importance of the investigation of disk substructures generated by models without planets. Although the majority of the substructures detected in recent high-res images of protoplanetary disks are being interpreted with models of disk-planet interaction \citep[e.g.,][]{Dipierro:2018,Zhang:2018}, this work shows how similar structures could be explained also via alternative scenarios involving gas instabilities and the interaction with the solid component in the disk. Quantitative investigations on the disk structures predicted by these models under reasonable conditions for the observed disks are therefore vital for an accurate interpretation of the recent results from high-res observations of disks.

\vskip 0.1in

\acknowledgements This work was
supported in part by the ngVLA Community Studies program, coordinated by the National Radio Astronomy Observatory,
which is a facility of the National Science Foundation operated
under cooperative agreement by Associated Universities, Inc.
D. B. acknowledges support from the Cal-Bridge scholarship program.

\end{document}